\documentclass[epj,referee]{svjour}
\usepackage{graphicx}
\begin{document}

\title{Elastic properties of a cellular dissipative structure}
\subtitle{Drift and oscillations in a 1-D pattern}
\author{P. Brunet \inst{1} \and J.-M. Flesselles \inst{1,2} \and L.
Limat \inst{1}}                     
%
%
\institute{\inst{1} Laboratoire PMMH-ESPCI - 10, rue
Vauquelin 75005 Paris France\\ \inst{2} Present Address: Saint-Gobain
Recherche - 39, quai Lucien Lefranc.  93303 Aubervilliers Cedex,
France}
\date{Received: date / Revised version: date}
%
\abstract{ Transition towards spatio-temporal chaos in one-dimensional
interfacial patterns often involves two degrees of freedom:
drift and out-of-phase oscillations of cells, respectively associated
to parity breaking and vacillating-breathing secondary bifurcations. 
In this paper, the interaction between these two modes is investigated
in the case of a single domain propagating along a circular array of
liquid jets.  As observed by Michalland and Rabaud for the printer's
instability \cite{Rabaud92}, the velocity $V_g$ of a constant width
domain is linked to the angular frequency $\omega$ of oscillations and
to the spacing between columns $\lambda_0$ by the relationship $ V_g =
\alpha \lambda_0 \omega$.  We show by a simple geometrical argument
that $\alpha$ should be close to $1/ \pi$ instead of the initial value
$\alpha = 1/2$ deduced from their analogy with phonons.  This fact is
in quantitative agreement with our data, with a slight deviation
increasing with flow rate.
} 
\maketitle
\section{Introduction}

Many studies have been devoted to the dynamics of one-dimensional
patterns.  One of the motivations is to seek for an equivalent of the
transition to turbulence in fluids (see \cite{Bohr} for a recent
exhaustive report).  Coullet and Iooss \cite{CoulletIooss90} have
classified the possible behavior in terms of secondary instabilities
linked to broken symmetries.  Most of them have been observed in
various experimental systems: Rayleigh-B\'enard convection
\cite{Dubois89}, directional solidification
\cite{Flesselles91,Faivre92,Gleeson91}, Taylor-Dean flow
\cite{Mutabazi93}, Taylor-Couette flow \cite{Wiener92}, directional
viscous fingering between two eccentric cylinders ("printer's
instability") \cite{Rabaud92,Pan93,Rabaud93}. These behaviors are
also recovered in numerical investigations of partial differential equations
governing interface instabilities \cite{ValanceMisbah94}, and also by
phenomenological models coupling a base mode k and its first harmonic 2k
\cite{CaroliFauve}.

\begin{figure}
(a)\includegraphics[width=0.38\textwidth]{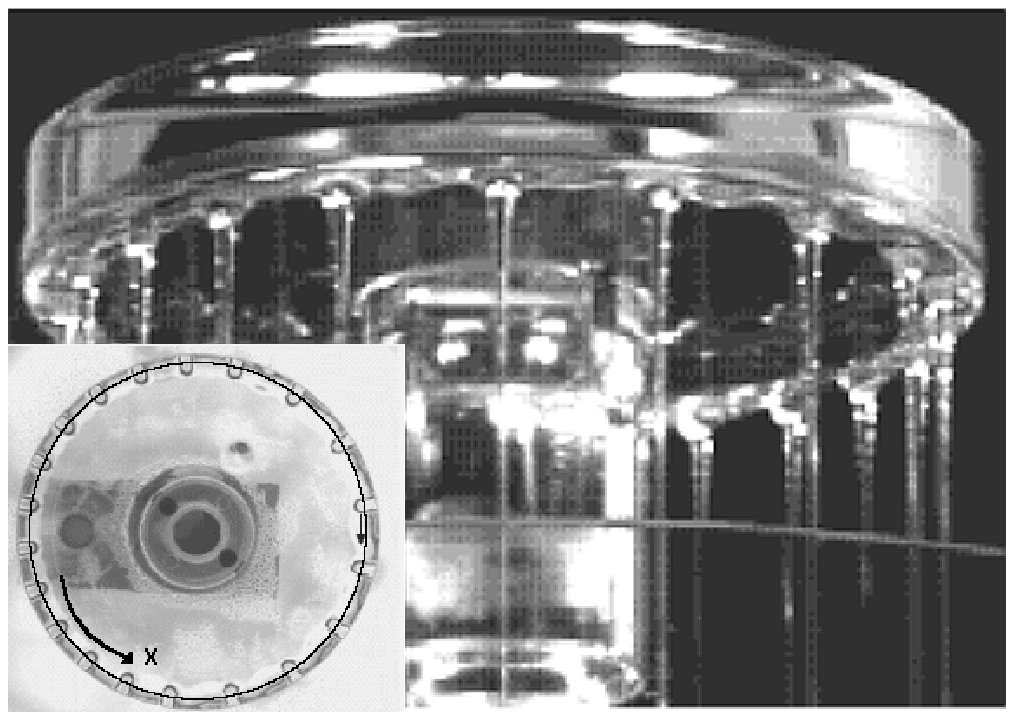}
(b)\includegraphics[width=0.38\textwidth]{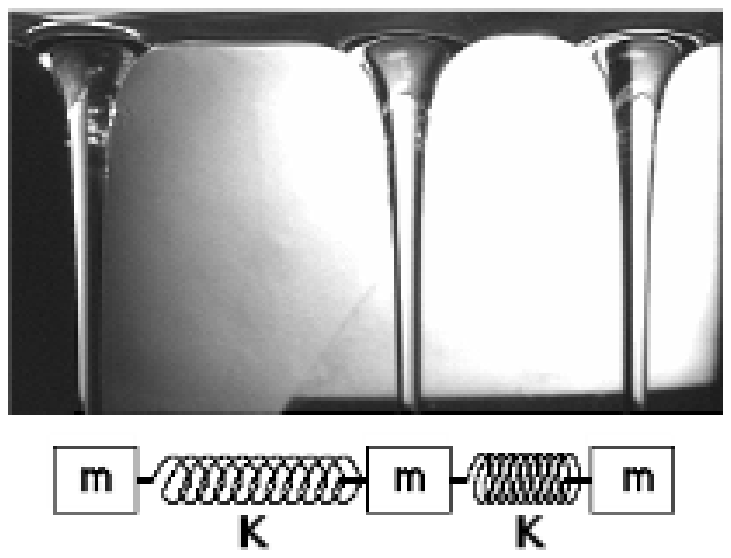}
\caption{(a) The circular fountain experiment : a pattern of liquid
columns is formed below an overflowing circular dish.  Insert : viewed
from above, columns appear as U-shaped spots.  (b) Like in the
printer's instability \cite{Rabaud92}, it is tempting to identify the
column array to a chain of springs and beads.}
\label{fig:dish}
\end{figure}

\begin{figure}
(a)\includegraphics[width=0.40\textwidth]{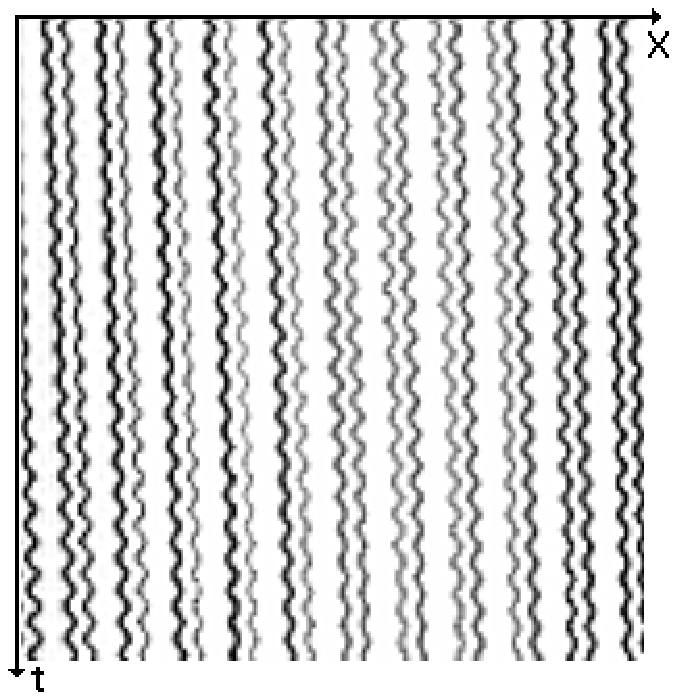}
(b)\includegraphics[width=0.40\textwidth]{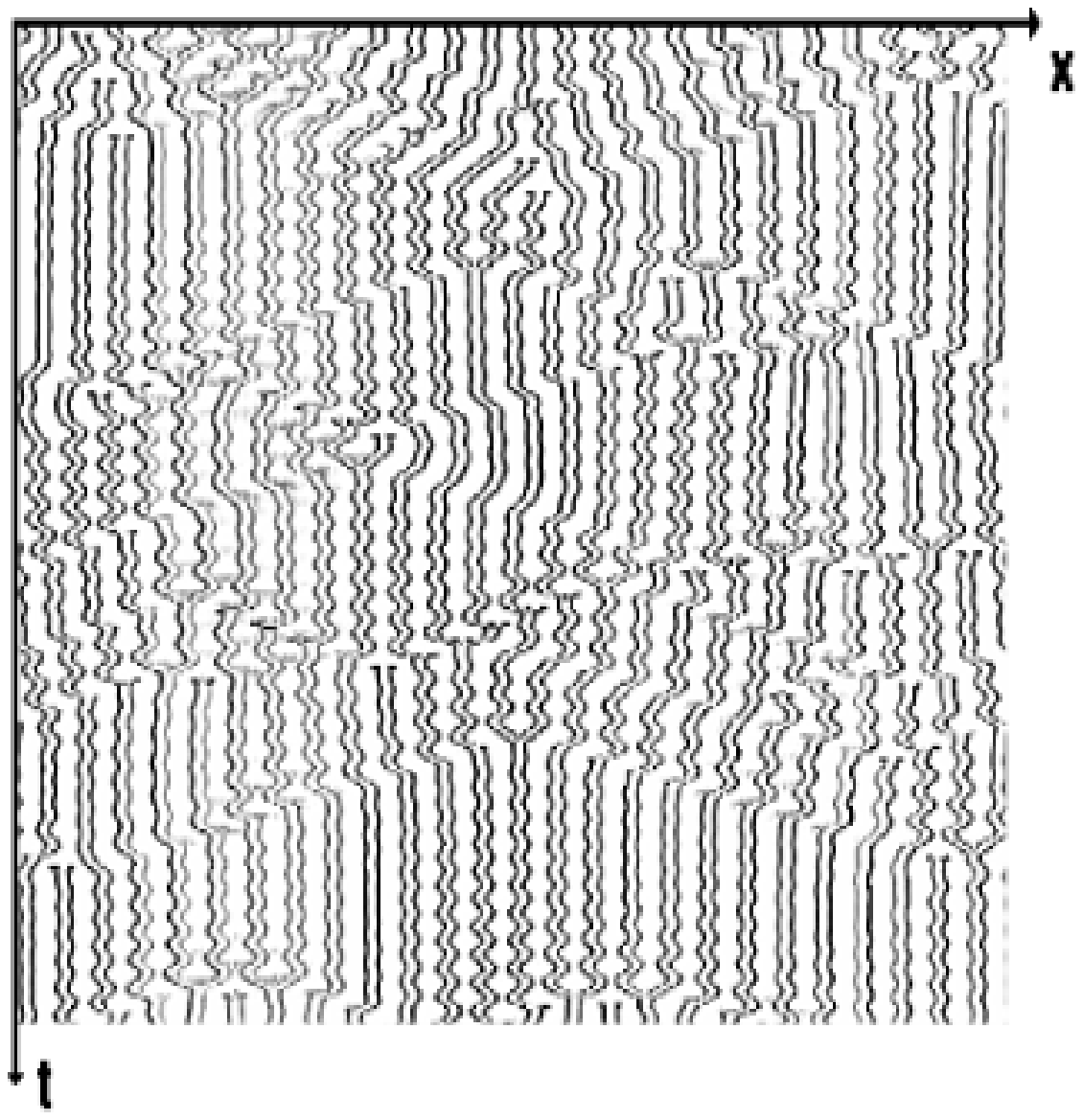}
\caption{Spatio-temporal diagrams (I).  (a) An oscillating regime (VB)
extended on the whole dish ($\Gamma$ = 0.31 cm$^2$/s, duration: $20$
s).  (b) Regime of spatio-temporal chaos.  Interactions between
oscillations and drift sustain disorder ($\Gamma$ = 0.39 cm$^2$/s).}
\label{fig:spatios}
\end{figure}

\begin{figure}
(a)\includegraphics[width=0.40\textwidth]{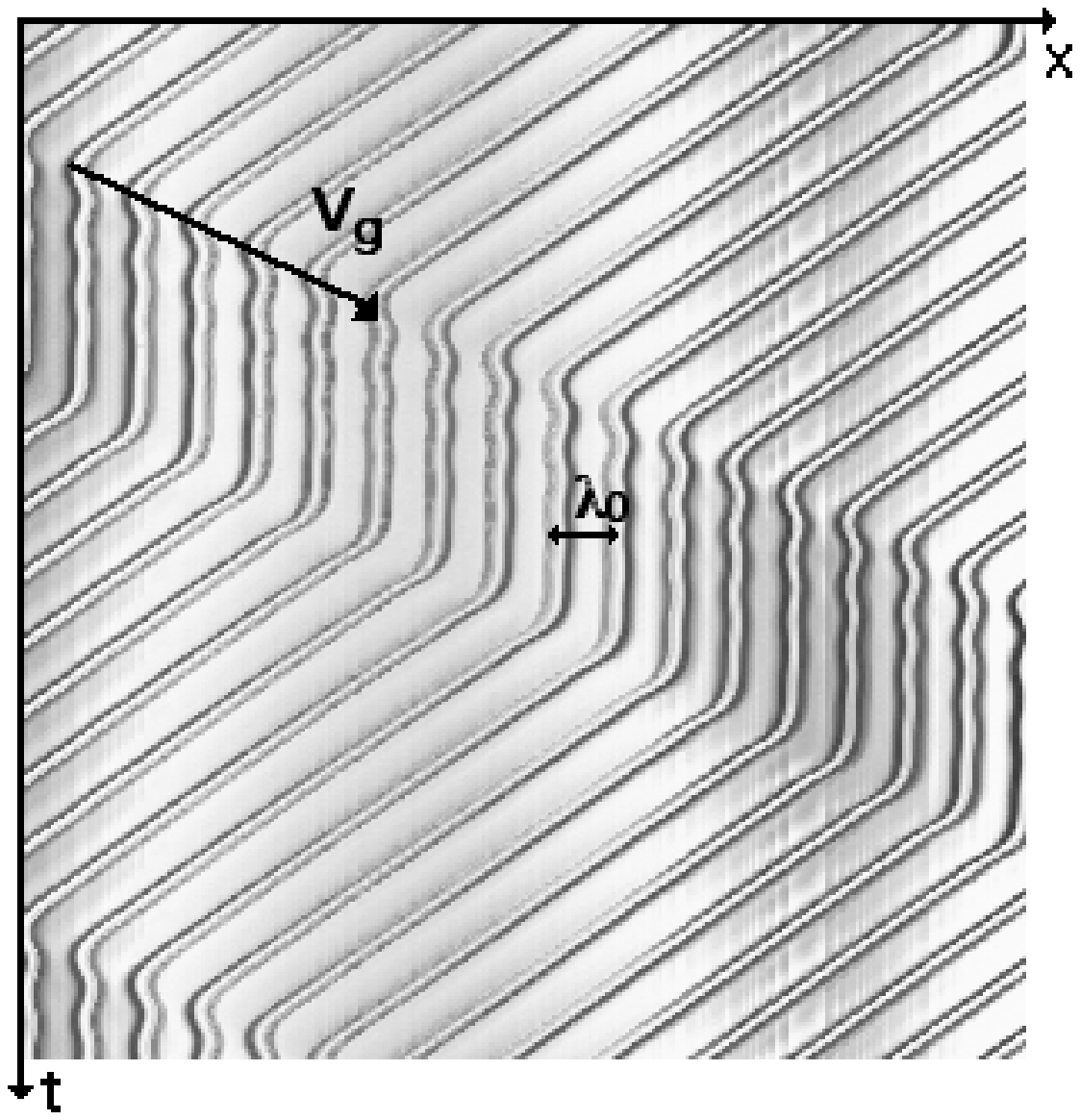}
(b)\includegraphics[width=0.40\textwidth]{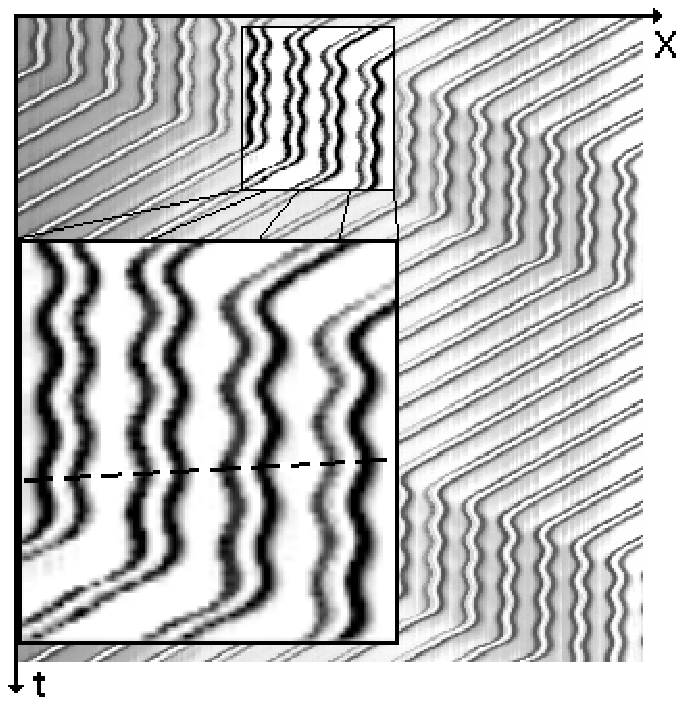}
\caption{Spatio-temporal diagrams (II). (a) and (b) Transient oscillations
following a dilation wave (PB mode)
(duration = $15$ s).  At low flow-rate ($\Gamma$ = 0.11 cm$^2$/s), the VB
mode is strongly damped (a).  At higher flow-rate($\Gamma$ = 0.29
cm$^2$/s), its lifetime reaches one period of rotation of the domain (b).
Inserted, a magnified view.  The dotted line linking the maxima of column
oscillations indicates the deviation to a perfect phase opposition between
neighbouring columns.}
\label{fig:spatios2}
\end{figure}

Our group has investigated another experimental system: a one
dimensional array of liquid columns formed below an overflowing
circular dish (see fig.  \ref{fig:dish}-a).  This system exhibits
dynamical behavior \cite{Limat98,Brunet01,Limat97} similar to those
observed in directional solidification or directional viscous
fingering.  Typical examples of spatio-temporal diagrams are displayed
on figs.  \ref{fig:spatios} and \ref{fig:spatios2}.  As in ref. 
\cite{Limat98,Brunet01}, these diagrams were obtained from pictures of
the dish taken from above (see insert of fig.  \ref{fig:dish}-a), and
by recording grey levels along a circle intersecting every column
trace.  Time runs from top to bottom, space (position along the dish
perimeter) is plotted on the horizontal axis.

Depending on the selected conditions (flow-rate, initial columns
number, possible initial imposed motions), different regimes occur,
the most spectacular one being the spatio-temporal chaos in fig. 
\ref{fig:spatios}-b.  This behavior involves a complex interaction
between two elementary degrees of freedom.  The first degree of
freedom is called a "vascillating-breathing" mode (fig. 
\ref{fig:spatios}-a): adjacent columns oscillate in phase-opposition,
which doubles the spatial period but preserves the $(x, -x)$ symmetry. 
The second degree of freedom is characterized by domains of asymmetric
cells that break the $(x, -x)$ symmetry.  This induces a drift of the
pattern \cite{Limat98,Brunet01,Limat97}.


These two bifurcations have motivated several experimental
\cite{Flesselles91,Faivre92,Rabaud93,Limat98,Brunet01,Limat97} as well
as theoretical studies
\cite{CoulletIooss90,ValanceMisbah94,Fauve91,CaroliFauve,C3G91,Gil99}. 
Among the latter, Misbah and Valance \cite{ValanceMisbah94} noticed
that for small systems, the interaction between these two modes may
lead to temporal chaos.  Therefore a careful examination of this
interaction in our well controled system is an essential step in the
study of the transition towards spatio-temporal chaos in an extended
geometry.  On the other hand, spatially chaotic states are so complex
that informations taken from regular behavior are to be prefered at
the present stage.  In various systems, including our fountain
experiment, the vacillating-breathing mode is known to accompany the
propagation of a solitary drifting domain, which trailing edge is
often followed by transient oscillations.  The interaction between a
solitary parity broken domain and its own oscillatory wake is thus the
subject of this article.

\section{Linking oscillation and drift: the phonon analogy}

The idea of a possible interaction between oscillations and drift in
1D patterns is natural, but remains undiscussed at least
quantitatively by available theories.  For instance, models based upon
the coupling between a base mode and its second harmonic ("k-2k
approaches") are able to capture both behavior analytically
\cite{Fauve91} but, to our knowledge, a solution involving a wall
separating oscillations and drift has never been proposed.  The other
well known approach based upon symmetry considerations
\cite{CoulletIooss90,C3G91} leads to sets of coupled phase and
amplitude equations that are specific to each state (oscillations and
drift) and are therefore unable to handle their interaction.  L. Gil
\cite{Gil99} recently showed that solitary drifting domains followed
by an oscillatory wake can be recovered when spatial phase shifts
between a base mode and a bifurcated oscillatory mode are considered. 
However, this study remained only qualitative.

On the other hand, quantitative evidences of such an interaction were
reported by Michalland and Rabaud for the printer's instability
\cite{Rabaud92} and confirmed by further investigations on the
fountain experiment \cite{Limat98,Limat97}.  These works all mention a
relationship linking the velocity of the domain walls $V_g$ to the
angular frequency $\omega$ of the transient oscillations left behind
and the spacing $\lambda_0$ between columns (defined outside of the
propagating domain):

\begin{equation}
V_g = \alpha \lambda_0 \omega
\label{eq:Vg}
\end{equation}

where $\alpha$ was found in a range from 0.36 to 0.4.  To interpret
qualitatively this result, Michalland and Rabaud \cite{Rabaud92}
suggested an analogy with the propagation of dilation waves on a
lattice of springs and beads (fig.~\ref{fig:dish}-b).  This "phonon
approach" may seem particularly relevant in the case of liquid columns
since this pattern appears as composed of discrete localized
singularities \cite{Limat97}.  The numerical simulation of spring and
mass lattice submitted to a sudden motion of a boundary indeed reveals
the propagation of dilation waves that exhibits a structure
qualitatively similar to that of our drifting domains, although they
are damped by dispersion on long time scales (fig. 
\ref{fig:spatios2}-a and b).  The velocity of these "domains" and the
angular frequency of oscillations are linked by a relationship
identical to (\ref{eq:Vg}), where $\lambda_0$ is the spacing between
each mass, but with $\alpha$ = $1/2$.

Though this was not explained in details in \cite{Rabaud92}, this
value of $1/2$ can be deduced from the dispersion relation of phonons
$\omega = 2 (K/m)^{1/2} \sin \vert ka/2 \vert$ governing the
propagation of waves on the lattice $x_n = \exp i(kna -\omega t) $
(where $k$ is the wavenumber, $a=\lambda_0$ the lattice spacing, $K$
the spring stiffness and $m$ the mass of each "atom").  Within this
framework, the domain velocity $V_g$ is identified to the group
velocity of phonons defined in the long wavelength limit ($k
\rightarrow 0$), while the frequency of oscillations is identified at
the boundary of the first Brillouin zone $(k=\pi/a)$, a limit which
corresponds to spatial-period doubling.

This approach has the merit to provide a simple interpretation of
(\ref{eq:Vg}).  However it is presumably incorrect, since $\alpha$ =
$1/2$ is inconsistent with the measured values (0.3 to 0.4).  As
recognized in \cite{Limat97}, including non-linearities and
dissipation in an improved "phonon" model does not solve this problem. 
This is puzzling because Michaland and Rabaud argument is rather
natural and very general.

Qualitatively, we think that a different coefficient is dictated by a
presumably non-linear effect: a phase-matching condition between
oscillation and drift.

\section{Geometrical argument}

A close examination of spatio-temporal diagrams (fig. 
\ref{fig:spatios2}) shows that there is a strong tendancy for phase
opposition between nearest neighbours in the transient oscillatory
wake.  The phonon model does not impose such a condition.  We think
this is the only reason for the discrepancy and postulate that columns
oscillate in perfect phase-opposition with nearest neighbours in the
oscillatory wake as usually assumed for a pure vascillating-breathing
mode \cite{CoulletIooss90,ValanceMisbah94}.  An idealized sketch of
the column motions involved at the rear of a drifting domain is
sugested on fig.  \ref{fig:geomrelation}-a .  In order to obtain a
stationary structure propagating uniformly with time, half a period of
oscillations $\pi / \omega$ must be equal to the time it takes to the
rear wall of the drifting domain to cover the spacing $\lambda_0$
between two oscillating columns $\lambda_0 / V_g$.  This geometrical
argument leads to:

\begin{equation}
V_g = \frac{1}{\pi} \lambda_0 \omega
\label{eq:unsurpi}
\end{equation}

and hence to $\alpha = \frac{1}{\pi}$.  Our argument can also be
stated as follows: since trajectories in the $(x,t)$ plane should be
continuous, $V_g$ must be equal (in absolute value) to the phase
velocity of oscillations $V_{\phi}=\omega / k$, where $k$, the
wavenumber of oscillations, is equal to $\pi / \lambda_0$ when
neighbouring cells oscillate in phase opposition.  A similar idea is
also perhaps underlying phase matching conditions invoked to describe
defect motions in cellular patterns \cite{AlvarezdeBruyn}.


\section{Experimental setup}

The experimental setup is simple : silicon oil (viscosity $\eta=100$
cP, surface tension $\gamma=20$ dyn/cm, density $\rho=0.97$ g/cm$^3$
at $20^o C$) overflows from a horizontal circular dish
(fig.~\ref{fig:dish}-a) at constant flow-rate per unit length
$\Gamma$.  For $\Gamma$ between 0.05 cm$^2$/s and 0.6 cm$^2$/s, the
liquid self-organizes as a pattern of liquid columns.  This pattern
results from a Rayleigh-Taylor instability combined with a constant
liquid supply.  It may bifurcate towards secondary instabilities,
which lead to the phenomenology displayed on figs.  \ref{fig:spatios}
and \ref{fig:spatios2}.

As explained above, these spatio-temporal diagrams are built from
pictures taken from above by a video camera, the fountain being
lighten by a circular neon tube (see insert of fig. 
\ref{fig:dish}-a)).  Columns then appear as U-shaped spots, moving
along a circle of radius $R=4.8$ cm slightly smaller than that of the
dish (5 cm here).  The diagrams are obtained by recording grey levels
along this circle.  Thanks to capillary effects, the columns can be
manipulated.  One can adjust their number and their initial motion
just by touching them with needles.  The rapid coalescence of several
columns with another one moving at constant speed induces a locally
heterogeneous pattern, in which a "dilation wave" followed by damped
oscillations (or more rigorously a drifting parity broken domain)
propagates around the dish.

As shown on \ref{fig:spatios2}-a and b, the length of the oscillatory
wake increases with flow-rate which makes possible accurate frequency
measurements.  We have systematically studied such domains, varying
the flow rate $\Gamma$ per unit length, and we have investigated the
evolution of their wall velocity $V_{g}$, as well as that of the
angular frequency $\omega$ and wavelength $\lambda_0$ left behind.

\section{Experimental results and discussion}


\begin{figure}
\includegraphics[width=0.32\textwidth]{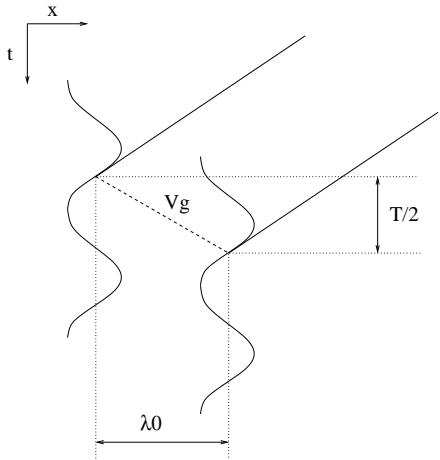}
\caption{Idealized geometry involved at the rear front of a
propagating parity broken domain, where the drift has to match with
transient oscillations.}
\label{fig:geomrelation}
\end{figure}

\begin{figure}
(a)\includegraphics[width=0.32\textwidth]{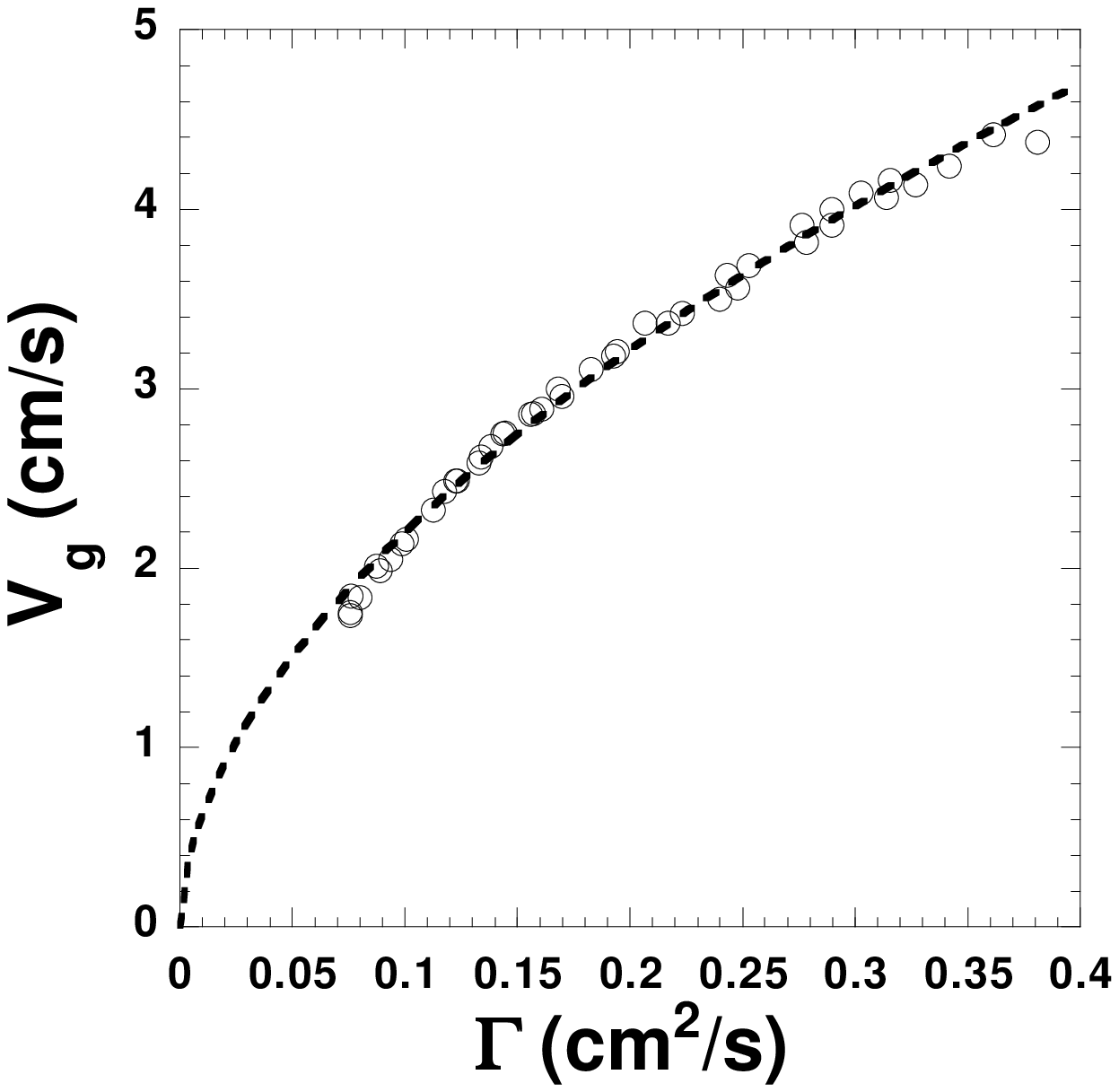}
(b)\includegraphics[width=0.32\textwidth]{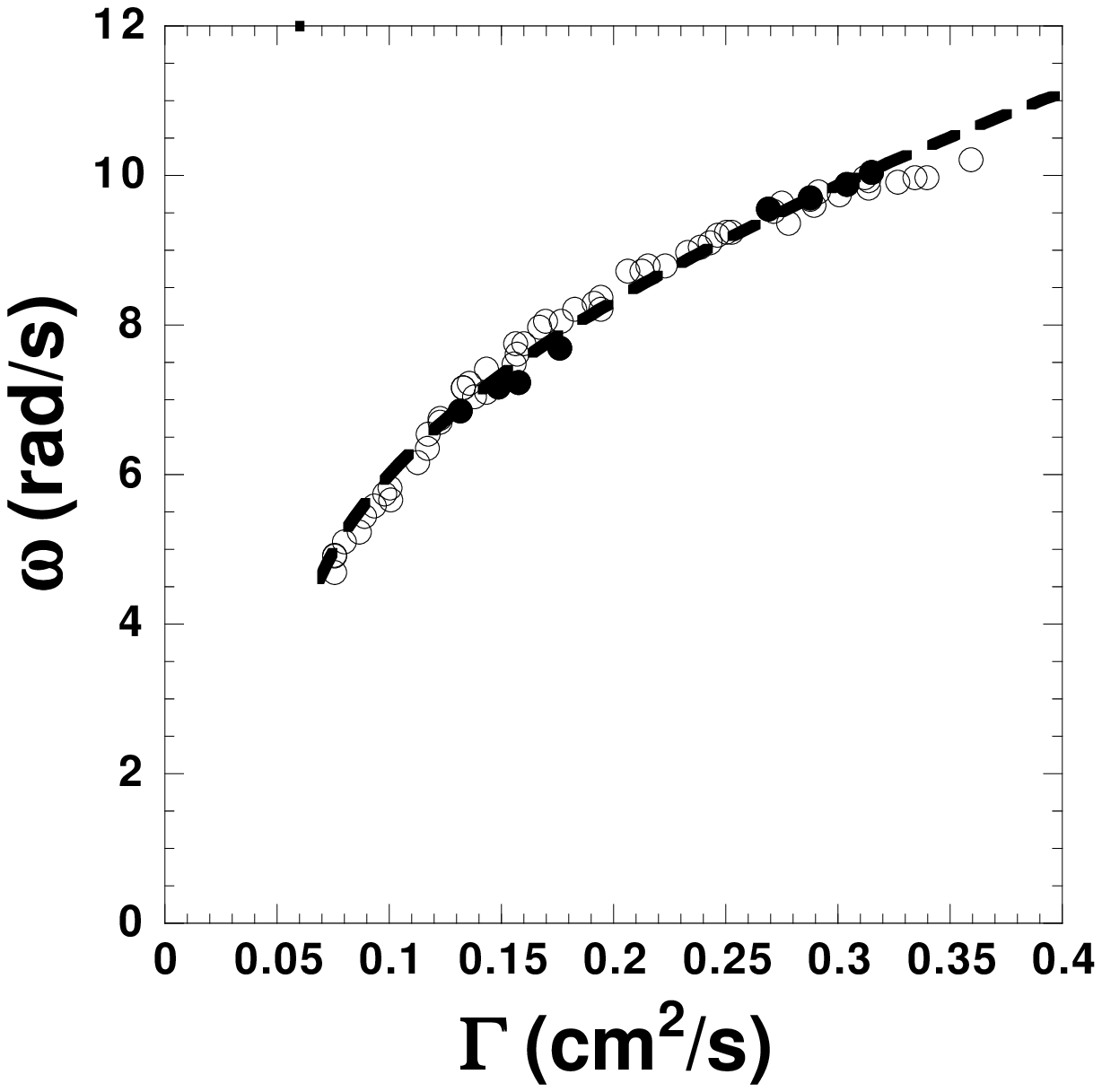}
(c)\includegraphics[width=0.32\textwidth]{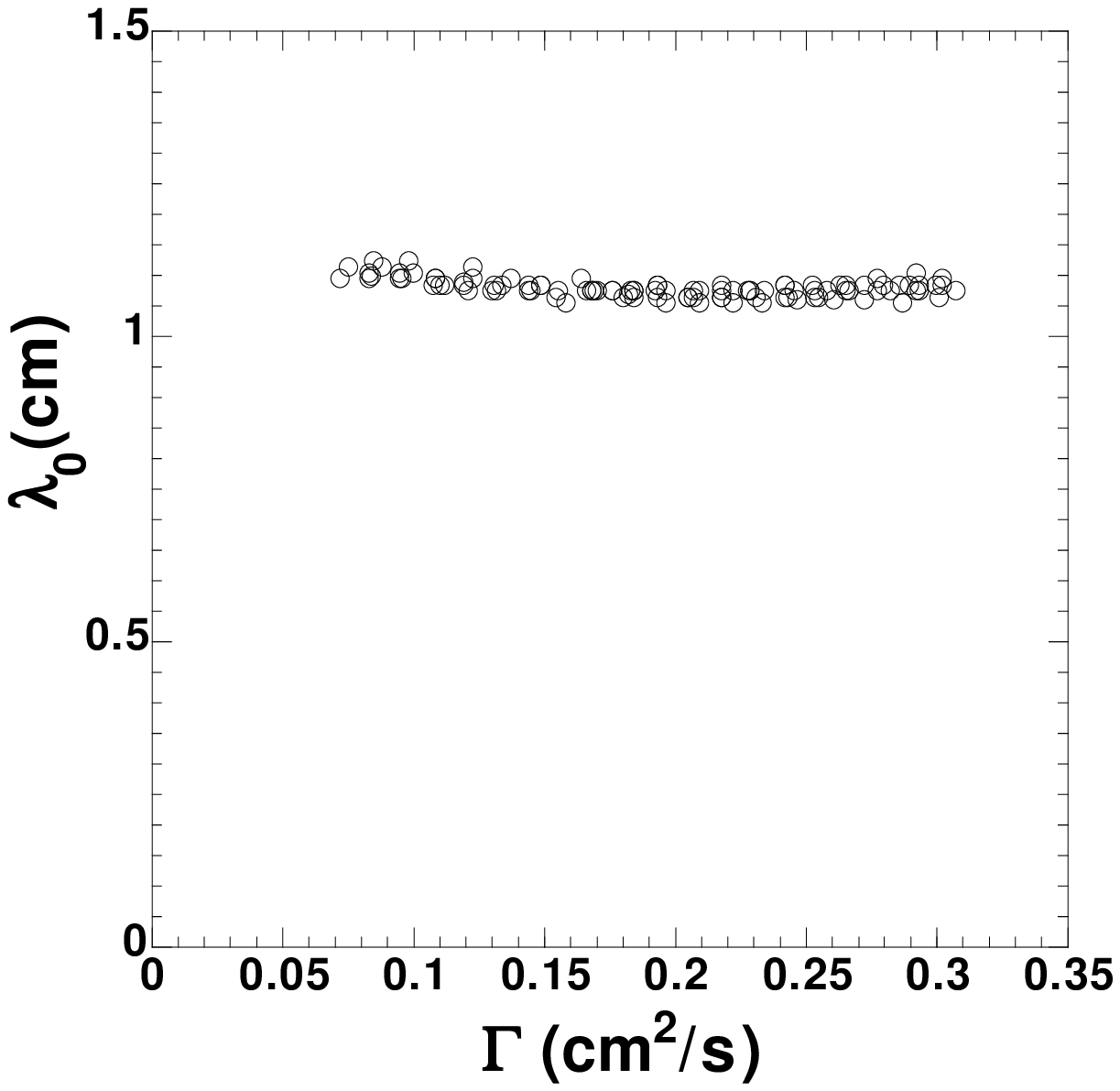}
\caption{(a) Measurements of domain walls $V_g$ versus flow-rate.  The
dotted line provides a fit with power law, with an exponant of close
to 0.5.  (b) Angular frequency of transient oscillations left behind a
drifting domain (open symbols) and in an extended oscillating state
(black symbols) versus flow-rate per unit length.  The fit traces a
square root law with a threshold.  (c) Wavelength $\lambda_0$ outside
a propagative domain, versus flow-rate, for several domain size.}
\label{fig:mesbrutes}
\end{figure}

The velocity of domain walls $V_g$ is plotted versus flow-rate on fig. 
\ref{fig:mesbrutes}-a.  Data are well fitted by a power law, with an
exposant close to 0.5, which suggests a relationship : $V_g \sim
\sqrt{\Gamma}$.

Measurements of the angular frequency $\omega$ of transient
oscillations left behind the propagating domains are plotted on
fig.~\ref{fig:mesbrutes}-b versus flow-rate per unit length $\Gamma$. 
At a few percent of accuracy, this frequency is very close to that of
global oscillations (black symbols) such as those on
fig.~\ref{fig:spatios}-a.  As in previous works
\cite{Limat98,Brunet01,Limat97}, $\omega$ increases with $\Gamma$.

Figure \ref{fig:mesbrutes}-c shows measurements of the wavelength
outside a propagative domain ($\lambda_0$).  This length does not vary
with flow-rate, neither with domain size.  Its value is around 1.08
cm.

Despite the obvious limitations of such an approach, it is here interesting
to confront this result to the "phonon analogy".  In this point of view,
assuming an effective column mass proportional to $\Gamma$, this
analogy would imply that the effective stiffness $K$ could scale as
$\Gamma^2$.  This dependance is compatible with an interaction between
columns through inertial terms of Navier-Stokes equation, the Reynolds
number $Re=\Gamma/\nu$ being here of order 1.

\begin{figure}
(a)\includegraphics[width=0.32\textwidth]{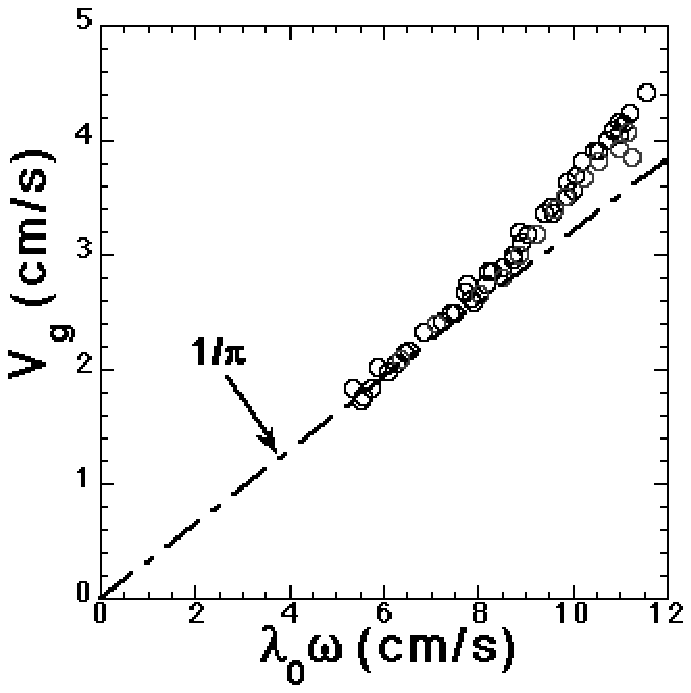}
(b)\includegraphics[width=0.32\textwidth]{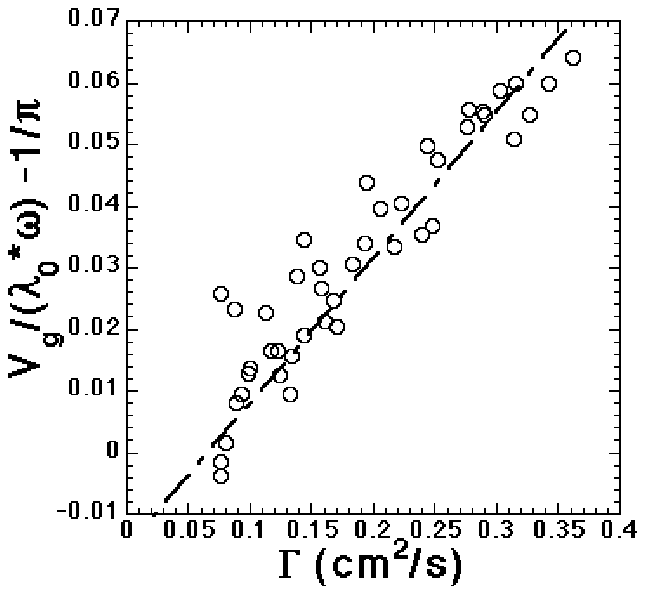}
(c)\includegraphics[width=0.32\textwidth]{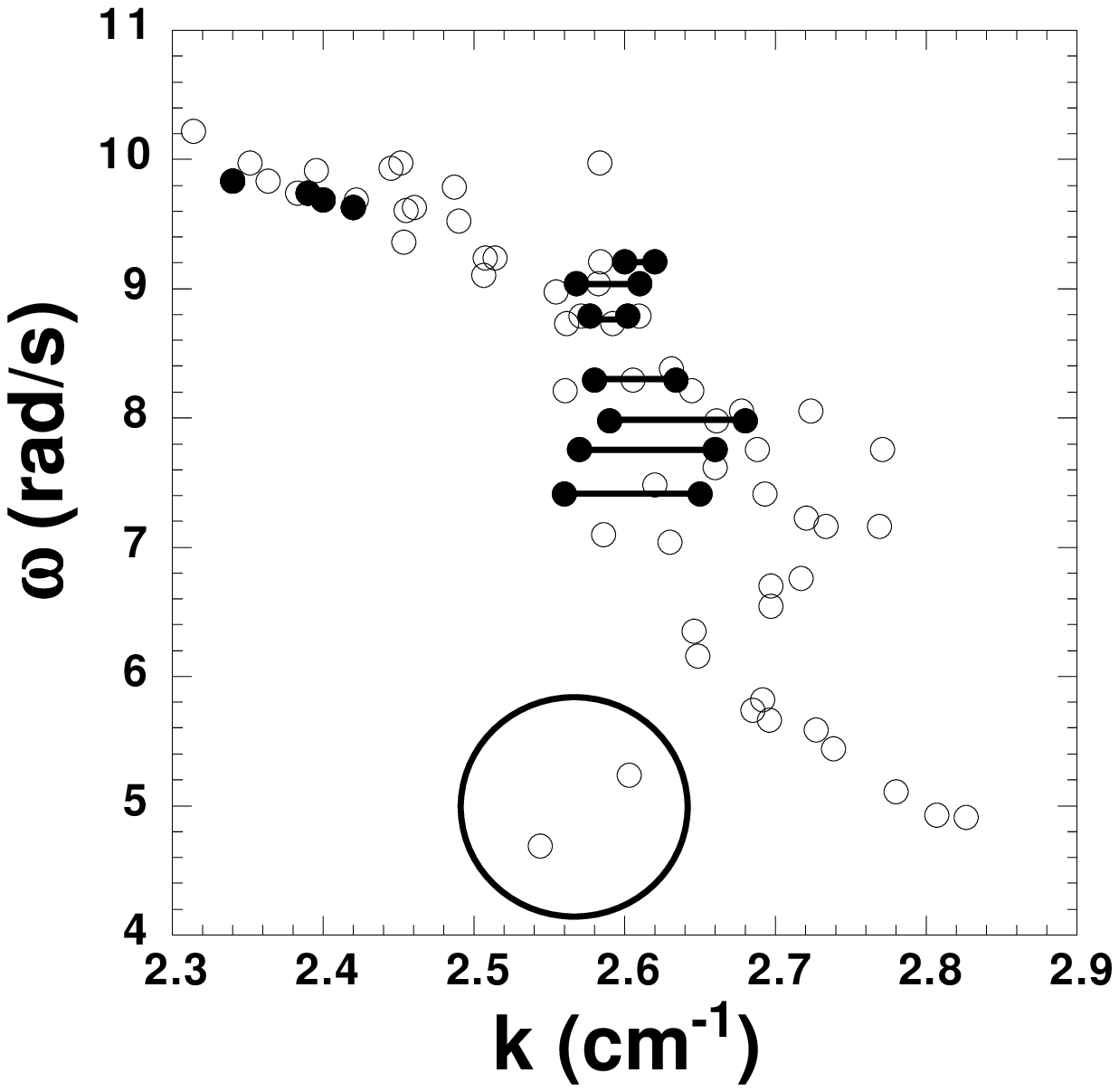}
\caption{(a) Velocity of propagative domains versus the quantity $
\lambda_0 \omega$ with $\omega=2\pi/T$.  (b) Difference between
$V_g/(\lambda_0 \omega) $ and $1/\pi$ versus flow-rate per unit
length.  (c) Evolution in the plane $(k,\omega)$ where $k$ is the
wavenumber of the representative points of the residual oscillations. 
Flow rate $\Gamma$ increases from right to left, from $\Gamma$ = 0.076
cm$^2$/s for $k$=2.83 cm$^{-1}$, to $\Gamma$ = 0.36 cm$^2$/s for
$k$=2.31 cm$^{-1}$.  Black circles are based upon direct $k$
measurements from spatio-temporal diagrams, while open symbols stand
for values deduced from $k = \omega / V_g$.  The two points inside the
circle correspond to negative values of $ V_g / \lambda_0 \omega - 1 /
\pi $.}
\label{fig:graph}
\end{figure}

In fig.~\ref{fig:graph}-a, we have plotted the values of $V_g$
versus $\lambda_0 \omega $.  The slope appears to be very close to $1/\pi$,
at low flow-rates, in agreement with eq.  (\ref{eq:unsurpi}), and seems to
become slightly larger when the flow-rate increases.  In fact the situation
is more complicated.  As appears on fig.~\ref{fig:graph}-b, there is a
systematic difference between $V_g$ and $\lambda_0 \omega $, that increases
linearily with $\Gamma$, following the empirical law :

\begin{equation}
\frac{V_g}{\lambda_0 \omega} - \frac{1}{\pi} = \frac{\Gamma - 
\Gamma_c}{D}
\label{eq:emplaw}
\end{equation}

$D$ has the dimension of a diffusion coefficient.  Its value is around
4.6 cm$^2$/s.  The threshold $\Gamma_c$ is around 0.06 cm$^2$/s.  

The effective value defined by $\alpha_{eff} = V_g / \lambda_0 \omega$
varies from $0.31$ to $0.38$ linearly with flow-rate.  The "ideal"
situation $\alpha =1/\pi$ implied by the argument suggested on fig. 
\ref{fig:geomrelation}-a only occurs close to the limit flow-rate
$\Gamma_c = 0.06 \pm 0.01$ cm$^2$/s.  It is interesting to note that
this limit flow rate nearly coincides with the critical threshold of
appearance of oscillations (fig.  \ref{fig:mesbrutes}-b).  Indeed,
angular frequency measurements are correctly fitted by the following 
law:

\begin{equation}
\omega = \omega_0 + \omega_1 \sqrt{\frac{\Gamma- \Gamma_c}{D}}  
\label{eq:omega}
\end{equation}

This is consistent with a Hopf bifurcation.  If this really holds, it
would suggest that (\ref{eq:Vg}) only holds in the limit of a
vanishing oscillation amplitude, the deviation from $1/\pi$ being
intrinsically non-zero for any flow rate.

Finding $\alpha = 1 / \pi$ instead of $1/2$ shows that the liquid column
array and an equivalent spring and beads sytem are rather different in the
details although the analogy can perhaps be useful to get reasonable
scaling laws.  As we have explained, the value found for $\alpha$ seems to
be dictated by : (1) a strong tendancy for neighboring columns to oscillate
in perfect phase opposition and (2) a "coherence" condition between the
wall motions and the oscillations.  In terms of Fourier analysis the first
condition can be expressed by $k=\pi/\lambda_0$, i.e. the system lies at
the boundary of the first Brillouin zone.

The second condition reads $V_g=V_{\phi}=\omega / k$, i.e. the velocity of
the domain walls must be equal to the phase velocity of the oscillations.
The first condition is rather reasonable if we think in terms of secondary
bifurcations of a dissipative pattern.  Near the threshold of oscillations,
we can speculate that a single wavevector and a single frequency are
selected.  In some sense this mode is the sole excited eigenmode of the
problem, whereas in a lattice of springs and beads, any eigenmode involved
in the dispersion relation can contribute to the oscillatory wake following
a dilation wave.  Since $\alpha$ differs from $1/\pi$ with increasing
flow-rate means that at least one of these two conditions is progressively
relaxed.  A careful examination of the spatio-temporal diagrams has
convinced us that only the first condition is violated, the second one
holding at any value of the flow rate.

This is clearly visible on the insert of fig.~\ref{fig:spatios2}-b.
Isophase lines linking second neighbours are slightly inclined, while the
structure of the trajectories remains consistent with the qualitative
picture of fig.  \ref{fig:geomrelation}-a.  To be more quantitative, we
have plotted on \ref{fig:graph}-c the path followed by the system in the
plane ($k$, $\omega$) when $\Gamma$ is increased.  Here we recall that
$\omega$ designates the frequency of the transient oscillations, while $k$
is their wave vector (which is different from the wave vector of the
pattern $k_0$), so that locally the column position varies as
$\exp[i(kx-\omega t)]$.  The black symbols are obtained by direct
determinations of $k$ on spatio-temporal diagrams across slope measurements
of the isophase line linking second neighbors (dotted line on
fig.~\ref{fig:spatios2}-a).  This slope $\psi$ reads:

\begin{equation}
\tan \psi = \frac{1}{\omega}(k - \pi /a)
\label{eq:mesk}
\end{equation}

The open symbols are obtained by using the relationship
$k=\omega/V_{\phi}$, and assuming $V_{\phi}=V_g$.  The fact that both
kind of symbols overlap shows that for any value of $\Gamma$, the
relationship $V_{\phi}=V_g$ holds.  Since the flowrate $\Gamma$
increases from right to left of this graph, the wave number $k$ of
perturbations of columns positions is equal to $\frac{\pi}{\lambda_0}$
= 2.9 cm$^{-1}$ at low flow rate and progressively decreases when
$\Gamma$ increases.  In other words, the system initialy lies at the
boundary of the Brillouin zone, and as $\Gamma$ increases, $k$ becomes
smaller than $\pi / \lambda_0$, the perfect phase opposition is
progressively lost.

\section{Conclusions - Conjectures}

In summary, this paper reports accurate measurements which allow us to
evidence a fundamental relationship linking the velocity of a
propagating parity broken domain to the frequency of its oscillatory
wake.  At a few percent of accuracy , this pulsation coincides with
that of the oscillatory mode itself observed alone in an extended
geometry (fig.~\ref{fig:mesbrutes}-b), which suggests that it is finaly
this oscillatory mode that rules the propagation of parity broken
domain.  Similarities and differences with phonons on a lattice of
springs and beads have been discussed in the plane $(k,\omega)$, the
residual oscillations exhibiting a possible shift of the wavenumber
$k$ with respect to the boundary of Brillouin zone of order of 20 $\%
$ of the maximal $k$ value.

We believe that the problem adressed in this article is important for
several reasons.  First, models based upon symmetry arguments
\cite{CoulletIooss90} miss the relationship reproduced in eq.
(\ref{eq:unsurpi}).  In this approach, $\omega$ is just a free parameter
that is selected at will to rebuild the spatio-temporal diagrams starting
from the amplitude evolutions.  This suggests that improvements of this
approach must be built \cite{Gil99}.  In another direction, it is to note
that k-2k models \cite{Fauve91,CaroliFauve} are able to capture both
oscillations and drift.  Therefore, a possible promising other idea to
interpret our result would consist in building a model of the wall
separating oscillations and drift by starting from Caroli et al equations
suggested in ref.\cite{CaroliFauve}.  This could constitute a better
framework to recover eq.\ref{eq:Vg} by a rigorous calculation.

Next, the point investigated here has something to do with an
insufficiently studied problem, i.e. that of Galilean invariance in pattern
dynamics.  For instance, Coullet and Fauve \cite{CoulletFauve85} studied
the effect of this invariance on a Ginzburg-Landau-like equation and
discussed the consequences for systems in which this invariance is slightly
broken by rigid boundary conditions. Another paper from Shraiman
\cite{Shraiman86} is also available in the case of the Kuramoto-Shivashinsky
equation.  Both papers show that, in such systems, a phase dynamics of
second order in time should be observed.  This second order time dynamics
is an alternative to the idea to invoke a column or cell inertia.  Finaly,
this subtle interaction between oscillations and drift is certainly
important in the genesis of spatio-temporal chaos, because it can influence
the nucleation process of defects, via "shock" formations in the phase
field.  We hope that our paper will motivate further studies in this field.


\begin{thebibliography}{}
%
%
\bibitem{Rabaud92}
S. Michalland \and M. Rabaud,
Physica D 61(1991) 197.

\bibitem{Bohr}
T. Bohr, M.H. Jensen, G. Paladin \and A. Vulpiani,
Dynamical systems approach to turbulence (Cambridge University
Press, 1998).

\bibitem{CoulletIooss90}
P. Coullet \and G. Iooss,
Phys. Rev. Lett. 64 (1990) 866.

\bibitem{Dubois89}
F. Daviaud, M. Dubois \and P. Berg\'e,
Europhys. Lett. 9 (1989) 441.

\bibitem{Flesselles91}
J.-M. Flesselles, A.J. Simon \and A.J. Libchaber,
Adv. Phys. 40 (1991) 1.

\bibitem{Faivre92}
G. Faivre \and J. Mergy,
Phys. Rev. A 45 (1992) 7320.

\bibitem{Gleeson91}
J.T. Gleeson, P.L. Finn \and P.E. Cladis
Phys. Rev. Lett. 66 (1991) 236.

\bibitem{Mutabazi93}
I. Mutabazi \and C.D. Andereck
Phys. Rev. Lett. 70 (1993) 1429.

\bibitem{Wiener92}
R. Wiener \and D.F. MacAlister
Phys. Rev. Lett. 69 (1992) 2915.

\bibitem{Pan93}
L. Pan \and J.R. de Bruyn
Phys. Rev. Lett. 70 (1993) 1791.

\bibitem{Rabaud93}
H.Z. Cummins, L. Fourtune \and M. Rabaud,
Phys. Rev. E 47 (1993) 1727.

\bibitem{ValanceMisbah94}
C. Misbah \and A. Valance
Phys. Rev. E 49 (1994) 166.

\bibitem{Limat98}
C. Counillon, L. Daudet, T. Podgorski \and L. Limat,
Phys. Rev. Lett. 80 (1998) 2117.

\bibitem{Brunet01}
P. Brunet, J.-M. Flesselles \and L. Limat,
Europhys. Lett. 56 (2001) 221.

\bibitem{Limat97}
F. Giorgiutti \and L. Limat,
Physica D 103 (1997) 590.

\bibitem{Fauve91}
S.Fauve, S. Douady \and O. Thual,
J. Phys. (Paris) 1 (1991) 311.

\bibitem{CaroliFauve}
B. Caroli, C. Caroli \and S. Fauve,
J. Phys. (Paris) 2 (1992) 281.

\bibitem{C3G91}
R.E. Goldstein, G.H. Gunaratne, L. Gil \and P. Coullet,
Phys. Rev. A 43 (1991) 6700.

\bibitem{Gil99}
L. Gil,
Europhys. Lett. 48 (1999) 156.


\bibitem{AlvarezdeBruyn}
R. Alvarez, M. van Hecke and W. van Saarloos, Phys. Rev. E, 56 (1997) R1306.
See also P. Habdas, M.J. Case and J.R. de Bruyn, Phys. Rev. E 63
(2001) 066305

\bibitem{Shraiman86}
B.I. Shraiman,
Phys. Rev. Lett. 57 (1986) 325.

\bibitem{CoulletFauve85}
P. Coullet \and S. Fauve,
Phys. Rev. Lett. 55 (1985) 2857.


\end{thebibliography}
%

\end{document}